\documentclass[aps,preprint,showpacs,preprintnumbers,amsmath,amssymb]{revtex4}

\usepackage{dcolumn}
\usepackage{bm}
\usepackage{amsmath,amssymb,epsfig,float}

\begin{document}

\title{Entanglement of coupled massive scalar
field in background of dilaton black hole}

\author{Jieci Wang, Qiyuan Pan, Songbai Chen and
Jiliang Jing\footnote{Corresponding author,
Email: jljing@hunnu.edu.cn}} \affiliation{ Institute of Physics and
Department of Physics,
\\ Hunan Normal University, Changsha, \\ Hunan 410081, P. R. China
\\ and
\\ Key Laboratory of Low-dimensional Quantum Structures
\\ and Quantum
Control of Ministry of Education, \\ Hunan Normal University,
Changsha, Hunan 410081, P. R. China}

\vspace*{0.2cm}
\begin{abstract}
\vspace*{0.2cm} The entanglement of the coupled massive scalar field
in the spacetime of a Garfinkle-Horowitz-Strominger(GHS) dilaton
black hole has been investigated. It is found that the entanglement
does not depend on the mass of the particle and the coupling between
the scalar field and the gravitational field, but it decreases as
the  dilaton parameter $D$ increases. It is interesting to note that
in the limit of $D\rightarrow M$, corresponding to the case of an
extreme black hole, the state has no longer distillable entanglement
for any state parameter $\alpha$, but the mutual information equals
to a nonvanishing minimum value, which indicates that the total
correlations consist of classical correlations plus bound
entanglement in this limit.
\end{abstract}

\vspace*{1.5cm}

\maketitle

\section{introduction}
Quantum entanglement is both the central concept and the major
resource in quantum information tasks such as quantum teleportation
and quantum computation \cite{Peres,Boschi,Bouwmeester,Pan}. As
relativistic field theory provides not only a more complete
theoretical framework but also many experimental setups,
relativistic quantum information theory may become an essential
theory in the near future with possible applications to quantum
entanglement and quantum teleportation. Thus, considerable effort
has been expended on the investigation of quantum entanglement in
the relativistic framework
\cite{Peres,Alsing-Milburn,Ge-Kim,Ling,Adesso}. It has been shown
that for scalar and Dirac fields, the degradation of entanglement
will occur from the perspective of a uniformly accelerated observer,
which essentially originates from the fact that the event horizon
appears and Unruh effect results in a loss of information for the
non-inertial observer
\cite{Schuller-Mann,Alsing-Mann,unruh,Qiyuan,Pan Qiyuan}.

On the other hand, string theory with an extra space compactified at
a larger length scale or lower energy scale than the Planck scale
has been an attractive idea to solve the gauge hierarchy problem and
possibly a candidate for quantum gravity \cite{arkani}. There is
also a growing interest in dilaton black holes from the string
theory in the last few years. Meanwhile, it is generally believed
that the study of quantum entanglement in the background of a
dilaton black hole may lead to a deeper understanding of black holes
and quantum gravity because it is related to the quantum information
theory, string theory and loop quantum gravity \cite{Dreyer,chen}.
In this paper, we will analyze the entanglement for the coupled
massive scalar field in the spacetime of a GHS dilaton black hole,
which was derived from the string theory. In particular, we here
choose the generically entangled state
$\sqrt{1-\alpha^{2}}|0\rangle_{A}|1\rangle_{B}
+\alpha|1\rangle_{A}|0\rangle_{B}$ rather than the maximally
entangled state
$\frac{1}{\sqrt{2}}(|0\rangle_{A}|0\rangle_{B}+|1\rangle_{A}|1\rangle_{B})$
in an inertial reference frame. It seems to be an interesting study
to consider the influences of the dilaton of the black hole, the
mass of the particle and the coupling between the scalar field and
the gravitational field on the quantum entangled states and show how
they will change the properties of the entanglement. We assume that
Alice has a detector which only detects mode $|n\rangle_{A}$ and Bob
has a detector sensitive only to mode $|n\rangle_{B}$, and they
share a generically entangled state at the same initial point in
flat Minkowski spacetime before the black hole is formed. After the
coincidence of Alice and Bob, Alice stays stationary at the
asymptotically flat region, while the other observer, Bob, moves
from the flat place toward the dilaton black hole. This won't change
the metric outside of the black hole and therefore won't change
Bob's acceleration \cite{Birkhoff}. Thus, Bob's detector registers
only thermally excited particles due to the Hawking effect
\cite{unruh-1}.

The outline of this paper is as follows. In Sec. 2 we discuss vacuum
structure of coupled massive scalar field in the spacetime. In Sec.
3 we analyze the effects of the dilaton parameter $D$, mass of the
particle and the coupling between the scalar field and the
gravitational field on the entanglement between the modes for the
different state parameter $\alpha$. We summarize and discuss our
conclusions in the last section.

\section{Vacuum structure of coupled massive scalar field}

The metric for a GHS black hole spacetime can be expressed as
 \cite{Horowitz}
 \begin{eqnarray}
 ds^2=-\left(\frac{r-2M}{r-2D}\right)dt^2+\left(\frac{r-2M}{r-2D}\right)^{-1}
 dr^2+r(r-2D)d\Omega^2,\label{gem1}
 \end{eqnarray}
where $M$ and $D$ are parameters related to mass of the black hole
and dilaton field. The relationship among $M$, the charge $Q$ and
$D$ is described as $D=Q^2/2M$. Throughout this paper we use
$G=c=\hbar=\kappa_{B}=1$.

The general perturbation equation for a coupled massive scalar field
in this dilaton spacetime is given by \cite{chen}
 \begin{eqnarray}
 \frac{1}{\sqrt{-g}}\partial_\mu(\sqrt{-g}g^{\mu\nu}
 \partial_\nu)\psi-(\mu+\xi R)\psi=0,\label{eq1}
 \end{eqnarray}
where  $\mu$  is the mass of the particle, $\psi$  is the scalar
field and $R$ is the Ricci scalar curvature. The coupling between
the scalar field and the gravitational field is represented by the
term $\xi R\psi$, where $\xi$ is a numerical coupling factor. After
expressing the normal mode solution as
\begin{eqnarray}
\psi_{\omega lm}=\frac{1}{h(r)}\chi_{\omega
l}(r)Y_{lm}(\theta,\varphi)e^{-i\omega t},
\end{eqnarray}
where $Y_{lm}(\theta,\varphi)$ is a scalar spherical harmonic on the
unit twosphere and $h(r)=\sqrt{r(r-2D)}$, we can easily get the
radial equation
\begin{eqnarray}\label{radial equation}
\frac{d^{2}\chi_{\omega
l}}{dr_{*}^{2}}+[\omega^{2}-V(r)]\chi_{\omega l}=0,
\end{eqnarray}
with
\begin{eqnarray}
V(r)=\frac{f(r)}{h(r)}\frac{d}{dr}\left[f(r)\frac{d
h(r)}{dr}\right]+\frac{f(r)l(l+1)}{h^{2}(r)}+f(r)\bigg[\mu^2+
\frac{2\xi D^{2}(r-2M)}{r^{2}(r-2D)^{3}}\bigg],
\end{eqnarray}
where $dr_{*}=dr/f(r)$ is the tortoise coordinates and
$f(r)=(r-2M)/(r-2D)$.

Solving Eq. (\ref{radial equation}) near the event horizon, we
obtain the incoming mode which is analytic everywhere in the
spacetime manifold
\begin{eqnarray}
\phi_{in,\omega lm}=e^{-i\omega v}Y_{lm}(\theta,\varphi),
\end{eqnarray}
and the outgoing mode for the inside and outside region of the event
horizon
\begin{eqnarray}\label{inside mode}
\phi_{out,\omega lm}(r<r_{+})=e^{i\omega u}Y_{lm}(\theta,\varphi),
\end{eqnarray}
\begin{eqnarray}\label{outside mode}
\phi_{out,\omega lm}(r>r_{+})=e^{-i\omega u}Y_{lm}(\theta,\varphi),
\end{eqnarray}
where $v=t+r_{*}$ and $u=t-r_{*}$. Eqs. (\ref{inside mode}) and
(\ref{outside mode}) are analytic inside and outside the event
horizon respectively, so they form a complete orthogonal family.

By defining the generalized light-like Kruskal coordinates
\cite{Ge-Kim}
\begin{eqnarray}
&&u=-4(M-D)\ln[-U/(4M-4D)],\quad \nonumber\\
&&v=4(M-D)\ln[V/(4M-4D)],\quad {\rm if\quad r>r_{+}};\nonumber\\
&&u=-4(M-D)\ln[U/(4M-4D)],\quad\nonumber\\
&&v=4(M-D)\ln[V/(4M-4D)], \quad {\rm if\quad r<r_{+}},
\end{eqnarray}
we can rewrite Eqs. (\ref{inside mode}) and (\ref{outside mode}) in
the following form
\begin{eqnarray}\label{inside mode1}
\phi_{out,\omega
lm}(r<r_{+})=e^{-4(M-D)i\omega\ln[-U/(4M-4D)]}Y_{lm}(\theta,\varphi),
\end{eqnarray}
\begin{eqnarray}\label{outside mode1}
\phi_{out,\omega
lm}(r>r_{+})=e^{4(M-D)i\omega\ln[U/(4M-4D)]}Y_{lm}(\theta,\varphi).
\end{eqnarray}
By using the formula $-1=e^{i\pi}$ and making (\ref{inside mode1})
analytic in the lower half-plane of $U$, we find a complete basis
for positive energy $U$ modes
\begin{eqnarray}\label{mode1}
\phi_{I,\omega lm}=e^{2\pi\omega (M-D)}\phi_{out,\omega
lm}(r>r_{+})+e^{-2\pi\omega (M-D)}\phi^{*}_{out,\omega lm}(r<r_{+}),
\end{eqnarray}
\begin{eqnarray}\label{mode2}
\phi_{II,\omega lm}=e^{-2\pi\omega (M-D)}\phi^{*}_{out,\omega
lm}(r>r_{+})+e^{2\pi\omega (M-D)}\phi_{out,\omega lm}(r<r_{+}).
\end{eqnarray}
Eqs. (\ref{mode1}) and (\ref{mode2}) are complete basis for positive
frequency modes which analytic for all real $U$ and $V$. Thus, we
can also quantize the quantum field in terms of $\phi_{I,\omega lm}$
and $\phi_{II,\omega lm}$ in the Kruskal spacetime.

Using the second-quantizing the field in the exterior of this
dilaton black hole \cite{Ge-Kim,unruh,Pan Qiyuan}, we can obtain the
Bogoliubov transformations for the particle annihilation and
creation operators in the dilaton and  Kruskal spacetime
\begin{eqnarray}
&&a_{K,\omega lm}=\frac{b_{out,\omega
lm}}{\sqrt{1-e^{-8\pi\omega (M-D)}}}-\frac{b^{\dag}_{in,\omega lm}}{\sqrt{e^{8\pi\omega (M-D)}-1}}, \nonumber\\
&&a^{\dag}_{K,\omega lm}=\frac{b^{\dag}_{out,\omega
lm}}{\sqrt{1-e^{-8\pi\omega (M-D)}}}-\frac{b_{in,\omega
lm}}{\sqrt{e^{8\pi\omega (M-D)}-1}},
\end{eqnarray}
where  $a_{K,\omega lm}$ and  $a^{\dag}_{K,\omega lm}$ are the
annihilation and creation operators acting on the Kruskal vacuum of
the exterior region, $b_{in,\omega lm}$ and $b^{\dag}_{in,\omega
lm}$ are the annihilation and creation operators acting on the
vacuum of the interior region of the black hole, and $b_{out,\omega
lm}$ and $b^{\dag}_{out,\omega lm}$ are the annihilation and
creation operators acting on the vacuum of the exterior region
respectively.

Now the Kruskal vacuum $|0\rangle_{K}$ outside the event horizon is
defined by
\begin{eqnarray}\label{Kruskal vacuum}
a_{K,\omega lm}|0\rangle_{K}=0.
\end{eqnarray}
After properly normalizing the state vector, we obtain the Kruskal
vacuum which is a maximally entangled two-mode squeezed state
\cite{Barnett,Ahn}
\begin{eqnarray}\label{Scalar-vacuum}
|0\rangle_{K}=\sqrt{1-e^{-8\pi\omega (M-D)}}
\sum_{n=0}^{\infty}e^{-4n\pi\omega
(M-D)}|n\rangle_{in}\otimes|n\rangle_{out},
\end{eqnarray}
and the first excited state
\begin{eqnarray}\label{Scalar-excited}
&&|1\rangle_{K}=a^{\dag}_{K,\omega lm}|0\rangle_{K}\nonumber\\
&&=[1-e^{-8\pi\omega (M-D)}]\sum_{n=0}^{\infty}
\sqrt{n+1}~e^{-4n\pi\omega
(M-D)}|n\rangle_{in}\otimes|n+1\rangle_{out},
\end{eqnarray}
where $\{|n\rangle_{in}\}$ and $\{|n\rangle_{out}\}$ are the
orthonormal bases for the inside and outside region of the event
horizon respectively. For the observer outside the black hole, he
needs to trace over the modes in the interior region since he has no
access to the information in this causally disconnected region.
Thus, the Hawking radiation spectrum can be obtained by
\begin{eqnarray}\label{Hawking}
N_\omega^{2}=_{K}\langle0|b^{\dag}_{K,\omega lm}b_{K,\omega
lm}|0\rangle_{K}=\frac{1}{e^{8\pi\omega (M-D)}-1},
\end{eqnarray}
Eq. (\ref{Hawking}) shows that the observer in the exterior of the
GHS dilaton black hole detects a thermal Bose-Einstein distribution
of particles as he traverses the Kruskal vacuum.

\section{Quantum entanglement in background of GHS dilaton black hole}

We will discuss quantum entanglement with the coupled massive scalar
field in the GHS dilaton black hole spacetime. We assume that Alice
has a detector which only detects mode $|n\rangle_{A}$ and Bob has a
detector sensitive only to mode $|n\rangle_{B}$, and they share a
generically entangled state at the same initial point in flat
Minkowski spacetime before the black hole is formed. The initial
entangled state is
\begin{eqnarray}\label{initial}
|\Psi\rangle=\sqrt{1-\alpha^{2}}|0\rangle_{A}|1\rangle_{B}
+\alpha|1\rangle_{A}|0\rangle_{B},
\end{eqnarray}
where $\alpha$ is some real number which satisfies
$|\alpha|\in(0,1)$, $\alpha$ and $\sqrt{1-\alpha^{2}}$ are the
so-called ``normalized partners". Using Eqs. (\ref{Scalar-vacuum})
and (\ref{Scalar-excited}), we can rewrite Eq. (\ref{initial}) in
terms of Minkowski modes for Alice and black hole modes for Bob.
Since Bob is causally disconnected from the interior region of the
black hole, we will take the trace over the states in this region
and obtain the mixed density matrix between Alice and Bob in the
exterior region
\begin{eqnarray}\label{Bos-density}
&&\rho_{AB}=[1-e^{-8\pi\omega
(M-D)}]\sum_{n=0}^{\infty}\rho_{n}~e^{-8n\pi\omega (M-D)},
\nonumber\\&& \rho_{n}=\alpha^{2}|1n\rangle\langle1n|
+(n+1)(1-\alpha^{2})[1-e^{-8\pi\omega
(M-D)}]|0(n+1)\rangle\langle0(n+1)|
\nonumber\\&&\qquad+\alpha\sqrt{(n+1)(1-\alpha^{2})
[1-e^{-8\pi\omega
(M-D)}]}|1n\rangle\langle0(n+1)|\nonumber\\&&\qquad+
\alpha\sqrt{(n+1)(1-\alpha^{2})[1-e^{-8\pi\omega (M-D)}]}
|0(n+1)\rangle\langle1n|,
\end{eqnarray}
where $|nm\rangle=|n\rangle_{A}|m\rangle_{B,out}$.

To determine whether this mixed state is entangled or not, we here
use the partial transpose criterion \cite{peres}. It states that if
the partial transposed density matrix of a system has at least one
negative eigenvalue, it must be entangled; but a state with positive
partial transpose can still be entangled. It is bound or
nondistillable entanglement. Interchanging Alice's qubits, we get
the partial transpose
\begin{eqnarray}\label{Bos-density1}
&&\rho_{AB}^{T_{A}}=[1-e^{-8\pi\omega
(M-D)}]\sum_{n=0}^{\infty}\rho_{n}^{\prime}~e^{-8n\pi\omega (M-D)},
\nonumber\\&& \rho_{n}^{\prime}=\alpha^{2}|1n\rangle\langle1n|
+(n+1)(1-\alpha^{2})[1-e^{-8\pi\omega
(M-D)}]|0(n+1)\rangle\langle0(n+1)|
\nonumber\\&&\qquad+\alpha\sqrt{(n+1)(1-\alpha^{2})
[1-e^{-8\pi\omega
(M-D)}]}|0n\rangle\langle1(n+1)|\nonumber\\&&\qquad+
\alpha\sqrt{(n+1)(1-\alpha^{2})[1-e^{-8\pi\omega (M-D)}]}
|1(n+1)\rangle\langle0n|,
\end{eqnarray}
and the corresponding negative eigenvalues of the partial transpose
in the ($n$,$n+1$) block is give by
\begin{eqnarray}
\lambda _{-}^{n}=\frac{e^{-8n\pi\omega (M-D)}[1-e^{-8\pi\omega
(M-D)}]}{2} \left[\beta_{n}-\sqrt{\beta_{n}^{2}
+4\alpha^{2}(1-\alpha^{2})[1-e^{-8\pi\omega (M-D)}]}~\right],
\end{eqnarray}
where $\beta_{n}=\alpha^{2}e^{-8\pi\omega
(M-D)}+n(1-\alpha^{2})[e^{-8\pi\omega (M-D)}-1]$. This mixed state
is always entangled for any finite value of $D$. The degree of
entanglement for the two observers here can be measured by using the
logarithmic negativity which serves as an upper bound on the
entanglement of distillation \cite{Vidal,Plenio}. This entanglement
monotone is defined as $N(\rho_{AB})=\log
_{2}||\rho_{AB}^{T_{A}}||$, where $||\rho_{AB}^{T_{A}}||$ is the
trace norm of the partial transpose $\rho_{AB}^{T_{A}}$. Thus, we
obtain the logarithmic negativity for this case
\begin{eqnarray}
N(\rho_{AB})=\log_{2}&\bigg\{&\alpha^{2}[1-e^{-8\pi\omega (M-D)}]
+\sum_{n=0}^{\infty}e^{-8n\pi\omega (M-D)}[1-e^{-8\pi\omega (M-D)}]
\nonumber\\
&&\times\sqrt{\beta_{n}^{2}
+4\alpha^{2}(1-\alpha^{2})[1-e^{-8\pi\omega (M-D)}]}\bigg\}
\end{eqnarray}
Note that the logarithmic negativity $N(\rho_{AB})$ is independent
of the mass of the particle $\mu$ and the numerical coupling factor
$\xi$. Thus, we can conclude that the mass of the particle and the
coupling between the scalar field and the gravitational field don't
influence the entanglement. But it is obvious that the dilaton
parameter $D$ has effect on the entanglement.

\begin{figure}[ht]
\includegraphics[scale=0.8]{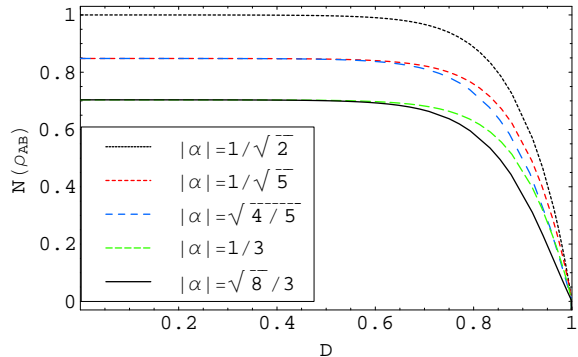}
\caption{\label{NegD}The logarithmic negativity as a function of the
dilaton parameter $D$ with the fixed $\omega$ and $M$ for different
$\alpha$.}
\end{figure}

The trajectories of the logarithmic negativity $N(\rho_{AB})$ versus
$D$ for different $\alpha$ in Fig. \ref{NegD} show how the dilaton
parameter $D$ would change the properties of the entanglement. The
logarithmic negativity $N(\rho_{AB})$ decreases as the dilaton
parameter $D$ increases, which shows that the monotonous decrease of
the entanglement with increasing $D$. It is interesting to note that
except for the maximally entangled state, the same ``initial
entanglement" for $\alpha$ and $\sqrt{1-\alpha^{2}}$ will be
degraded along two different trajectories, which just shows the
inequivalence of the quantization for a scalar field in the dilaton
black hole and Kruskal spacetimes. In the limit of $D\rightarrow M$,
corresponding to the case of an extreme black hole, the logarithmic
negativity is exactly zero for any $\alpha$, which indicates that
the state has no longer distillable entanglement. This is due to the
fact that the observer in the exterior of the GHS dilaton black hole
detects a thermal Bose-Einstein distribution of particles given by
Eqs. (\ref{Hawking}) as he traverses the Kruskal vacuum. This number
of the particles $N_\omega^{2}\rightarrow \infty$ in the limit of
$D\rightarrow M$, which means that the observer detected a maximally
mixed state which contains no information.

\begin{figure}[ht]
\includegraphics[scale=0.8]{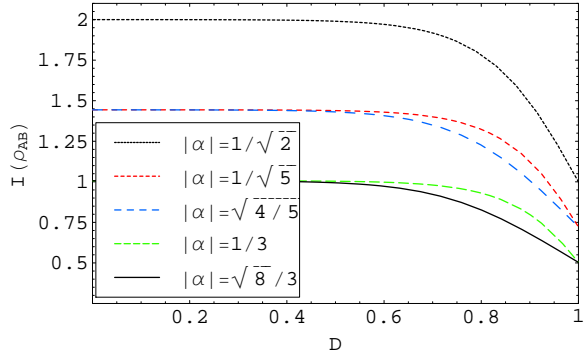}
\caption{\label{MuInD}The mutual information as a function of the
dilaton parameter $D$ with the fixed $\omega$ and $M$ for different
$\alpha$.}
\end{figure}

We may also estimate the total correlations between Alice and Bob by
use the mutual information \cite{RAM}
\begin{eqnarray}
I(\rho_{AB})=S(\rho_{A})+S(\rho_{B})-S(\rho_{AB}),
\end{eqnarray}
where $S(\rho )=-\text{Tr}(\rho \log_{2}\rho)$ is the entropy of the
density matrix $\rho$. The mutual information quantifies how much
information two correlated observers possess about one another's
state. The entropy of the joint state is
\begin{eqnarray}
&&S(\rho _{AB})=-\sum_{n=0}^{\infty } e^{-8n\pi\omega
(M-D)}[1-e^{-8\pi\omega (M-D)}]
\nonumber\\
&&\qquad\qquad~\times
\left\{\alpha^{2}+(n+1)(1-\alpha^{2})[1-e^{-8\pi\omega
(M-D)}]\right\} \log_{2}e^{-8n\pi\omega (M-D)}
\nonumber\\
&&\qquad\qquad~\times [1-e^{-8\pi\omega
(M-D)}]\left\{\alpha^{2}+(n+1)(1-\alpha^{2})[1-e^{-8\pi\omega
(M-D)}]\right\}.
\end{eqnarray}
 We obtain  Bob's entropy in exterior region of the event horizon by tracing over Alice's
  states for the density matrix $\rho_{AB}$
\begin{eqnarray}
&&S(\rho_{B})=-\sum_{n=0}^{\infty }e^{-8n\pi\omega
(M-D)}[1-e^{-8\pi\omega
(M-D)}]\left\{\alpha^{2}+n(1-\alpha^{2})[e^{8\pi\omega
(M-D)}-1]\right\}
\nonumber\\
&&\qquad\qquad\times \log_{2}e^{-8n\pi\omega (M-D)}[1-e^{-8\pi\omega
(M-D)}]\left\{\alpha^{2}+n(1-\alpha^{2})[e^{8\pi\omega
(M-D)}-1]\right\}.
\end{eqnarray}
Tracing over Bob's states, we can also find  Alice's entropy can be
expressed as
\begin{eqnarray}\label{Alice-entropy}
&&S(\rho_{A})=-[\alpha^{2}\log_{2}\alpha^{2}
+(1-\alpha^{2})\log_{2}(1-\alpha^{2})].
\end{eqnarray}
Thus, we draw the behaviors of the mutual information $I(\rho_{AB})$
as a function of the dilaton parameter $D$ for different values of
the state parameter $\alpha$ in Fig. \ref{MuInD}.

Fig. \ref{MuInD} shows that as the dilaton parameter $D$ increases,
the mutual information becomes smaller. Note that except for the
maximally entangled state, the same ``initially mutual information"
for $\alpha$ and $\sqrt{1-\alpha^{2}}$ will be degraded along two
different trajectories. In the limit of $D\rightarrow M$, the mutual
information converges to the same nonvanishing minimum value again.
Obviously if the ``initially mutual information" is higher, it is
degraded to a higher degree. Since the distillable entanglement in
the limit $D\rightarrow M$ is exactly zero for any $\alpha$, we can
say that the total correlations consist of classical correlations
plus bound entanglement in this limit.

It is interesting to compare the results of the GHS black hole with
those in the Schwarzschild one. For both the GHS and Schwarzschild
cases, when describing the state (which involves tracing over the
unaccessible modes), the observers find that some of the
correlations are lost \cite{Pan Qiyuan} due to the exterior region
is causally disconnected from the interior region of the black hole.
However, the entanglement is relevant to both the mass and dilaton
parameters of the black hole in the GHS case, but it depends only on
the mass of the black hole in the Schwarzschild case.

\section{summary}

We have analytically discussed the entanglement between two modes of
a coupled massive scalar field as detected by Alice who stays
stationary at an asymptotically flat region and Bob who locates near
the event horizon in the background of a GHS dilaton black hole. It
is shown that the entanglement does not depend on the mass of the
particle and the coupling between the scalar field and the
gravitational field, but it decreases with increasing dilaton
parameter $D$. It is found that the same ``initial entanglement" for
the state parameter $\alpha$ and its ``normalized partners"
$\sqrt{1-\alpha^{2}}$ will be degraded along two different
trajectories as the dilaton increases except for the maximally
entangled state $\alpha=1/\sqrt{2}$, which just shows the
inequivalence of the quantization for a scalar field in the dilaton
black hole and Kruskal spacetimes. In the limit of $D\rightarrow M$,
corresponding to the case of an extreme black hole, the state has no
longer distillable entanglement for any $\alpha$. However, further
analysis shows that the mutual information is degraded to a
nonvanishing minimum value in this limit, which indicates that the
total correlations consist of classical correlations plus bound
entanglement.

\begin{acknowledgments}
This work was supported by the National Natural Science Foundation
of China under Grant No.  10875040 and 10847124; the FANEDD under
Grant No. 200317; and the Hunan Provincial Natural Science
Foundation of China under Grant No. 08JJ3010.

\end{acknowledgments}

\end{document}